\documentclass[tighten,iop]{emulateapj}

\usepackage[english]{babel}
\usepackage{amsmath}
\usepackage{amsfonts}
\usepackage{amssymb}
\usepackage{graphicx}
\usepackage[dvips]{color} 

\usepackage{natbib}

\shorttitle{The Neutrino-Driven Jet Penetration}
\shortauthors{Nagakura}

\begin{document}

\title{The Propagation of Neutrino-Driven Jets in Wolf-Rayet Stars}

\author{Hiroki Nagakura$^{1,2}$}
\address{$^1$Yukawa Institute for Theoretical Physics, Kyoto
  University, Oiwake-cho, Kitashirakawa, Sakyo-ku, Kyoto, 606-8502,
  Japan}
\address{$^2$Advanced Research Institute for Science \&
Engineering, Waseda University, 3-4-1 Okubo,
Shinjuku, Tokyo 169-8555, Japan}

\begin{abstract}
We numerically investigate the jet propagation through a rotating collapsing Wolf-Rayet star with detailed central engine physics constructed based on the neutrino-driven collapsar model. The collapsing star determines the evolution of mass accretion rate, black hole mass and spin, all of which are important ingredients for determining the jet luminosity. We reveal that neutrino-driven jets in rapidly spinning Wolf-Rayet stars are capable of breaking out from the stellar envelope, while those propagating in slower rotating progenitors fail to jet breakout due to insufficient kinetic power. For progenitor models with successful jet breakouts, the kinetic energy accumulated in the cocoon could be as large as $\sim 10^{51}$erg and might significantly contribute to the luminosity of the afterglow emission or to the kinetic energy of the accompanying supernova if nickel production takes place. We further analyze the post breakout phase using a simple analytical prescription and conclude that the relativistic jet component could produce events with an isotropic-luminosity $L_{p(iso)}\sim 10^{52}$erg/s and isotropic-energy $E_{j(iso)}\sim 10^{54}$erg. Our findings support the idea of rapidly rotating Wolf-Rayet stars as plausible progenitors of GRBs, while slowly rotational ones could be responsible for low luminosity or failed GRBs.
\end{abstract}

\keywords{supernovae: general---neutrinos---hydrodynamics, black hole physics}

\section{Introduction}
Long-duration Gamma-Ray Bursts (GRBs) are thought to originate from the death of massive stars \citep{2006ARA&A..44..507W}. It is widely recognized that the study of GRBs provides the important knowledge on the final evolutionary stage in the life of massive stars. Although the nature of GRBs remains elusive, one viable scenario to produce a GRB is the neutrino-driven collapsar model \citep{1993ApJ...405..273W,1999ApJ...524..262M}. The gravitational collapse of the rapidly rotating core is believed to create a fast rotational Kerr black hole. The copious amounts of neutrinos and their anti particles, which are emitted from the hot accretion disk, annihilate and create an electron positron fireball around the rotational axis \citep{1989Natur.340..126E}. The baryon-starved fireball is believed to give rise to a relativistic collimated outflow, and eventually produce a GRB after the beam has broken free from the stellar progenitor.

Over the years, substantial work has been made towards understanding if neutrino-driven collapsar jets can produce the required relativistic outflow (see e.g. \citet{1999ApJ...524..262M,2003ApJ...588L..25F,2006ApJ...641..961L,2007ApJ...659..512N,2008ApJ...673L..43D,2009ApJ...692..804L,2010ApJ...720..614H,2011ApJ...737....6S,2011MNRAS.410.2385T}). However, there are still many open questions. In terms of the central engine, one of the largest uncertainties is the efficiency of energy deposition by neutrinos. Although numerical studies are very powerful methods to investigate the energy deposition rate by neutrinos and subsequent evolution of the jet, we need to solve General Relativistic Neutrino Radiation Hydrodynamics with microphysics for several tens of seconds after the black hole formation. This is certainly challenging as the typical life time of the central engine is roughly six orders of magnitude longer than the dynamical timescale of the nascent black hole. Computational resources are not yet available to perform such numerical studies. Because of these difficulties, a number of studies have thus far employed steady state approximations \citep{1999ApJ...518..356P,1994ApJ...428L..13N,2002ApJ...577..311K,2002ApJ...579..706D,2005ApJ...629..341K,2006ApJ...643L..87G,2007ApJ...657..383C,2010ApJ...709..851L,2011MNRAS.410.2302Z,2012Ap&SS.337..711L} or performed hydrodynamic (or magnetohydrodynamic) simulations \citep{2005ApJ...632..421L,2007ApJ...664.1011J,2007PThPh.118..257S,2010ApJ...720..614H} with simplified microphysical treatments. It is interesting to note that \citet{2011MNRAS.410.2302Z} recently have conducted General Relativistic Ray Tracing Neutrino Radiation Transfer and they found that the energy deposition by neutrinos could be well described as a simple analytic formula. Thanks to these studies, the jet luminosity can be estimated qualitatively without the need for expensive numerical simulations.

It should be noted, however, even if the neutrino-driven jet is successfully launched, this does not guarantee the production of a GRB. As a minimum requirement, the jet needs to successfully penetrate the stellar envelope otherwise it would become non-relativistic and thus incapable of producing GRB. Ever since the neutrino-driven collapsar model was proposed, a number of numerical works of jet propagation have been done (see e.g. \citep{2000ApJ...531L.119A,2003ApJ...586..356Z,2007ApJ...665..569M,2009ApJ...699.1261M,2011ApJ...731...80N}). In these studies, however, the jet luminosity was assumed for simplicity to either be constant or to directly follow the mass accretion rate (see e.g. \citet{2001ApJ...550..410M,2012ApJ...754...85N}). In order to judge whether neutrino-driven jets are capable of successfully breaking out of their progenitors, and to explore the effects of rotation, one needs to take into account how the jet power scales with the neutrino energy deposition generated by the accompanying accretion. In this paper, we present, for the first time, the propagation of neutrino-driven jets employing accurate neutrino energy deposition rate as calculated by \citet{2011MNRAS.410.2302Z}. The evolution of mass accretion rate and black hole mass and spin, all of which are necessary to evaluate the energy deposition by neutrinos, are evaluated here using an inner boundary condition in the simulation, although the current calculation is still not self-consistent as it fails to resolve the accretion disk and assumes no feedback (e.g. a disk wind). The purpose of this study is (1) to clarify whether neutrino-driven jets can successfully break out from the stellar surface and (2) to determine the progenitor's rotation rate necessary for successful jet breakout. As we will show in this paper, the final outcome of the explosion depends sensitive on the rate of rotation, and differences in rotation rate could be responsible for the observational differences seen between GRBs, low luminosity GRBs (LLGRBs) and failed GRBs.

\section{Methods and Models}

\begin{figure}
\vspace{15mm}
\epsscale{1.2}
\plotone{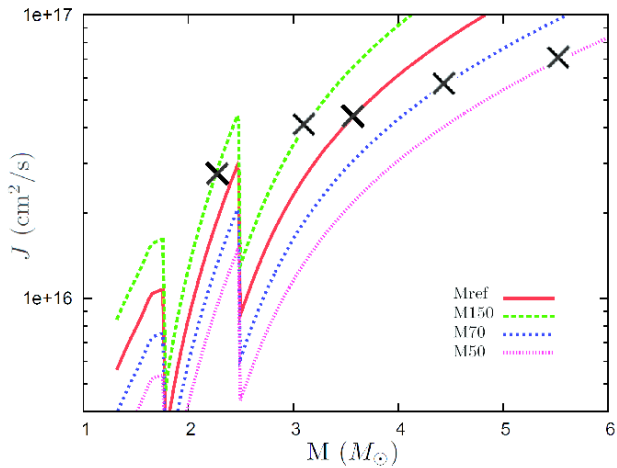}
\caption{SAM distribution of models (Red: Mref, Green: M150, Blue: M70, Pink: M50). The cross marks denote the intersection between the SAM of star and SAM at the last stable orbit for a black hole with the mass and angular momentum inside the indicated coordinate.
\label{f1}}
\end{figure}

We perform two dimensional, relativistic hydrodynamical axisymmetric (and also equatorial symmetric) simulations of the accretion and subsequent jet propagation. The numerical code employed in this paper is essentially the same as those used in previous papers \citep{2011ApJ...731...80N,2012ApJ...754...85N}. The initial stellar density distribution is fixed as 16TI model in \citet{2006ApJ...637..914W}. As is the case with previous studies, we cut the inner portions of the star from a certain radius. The self-gravity of matter in the active numerical regions is calculated by solving Poisson equations and the monopole gravity is added as the point mass at the inner excised region. The mass accretion rate ($\dot{M}$) is estimated by the mass flows through the inner boundary (see Eq.~(1) in \citet{2012ApJ...754...85N}). The mass and angular momentum in the excised region are assumed to be the same mass and angular momentum of the black hole. The time evolution of mass and spin of black hole is calculated by integrating mass and angular momentum flux crossing the inner boundary. It should be noted that when the specific angular momentum (SAM) at the location of inner boundary in equatorial region becomes larger than SAM at the inner most stable circular orbit (ISCO) (see cross marks in Figure~\ref{f1}), we alter our prescription when calculating the angular momentum to:
\begin{eqnarray}
f_a = \dot{M} \times J_{ISCO} \label{eq:anguinteg},
\end{eqnarray}
where $f_a$ and $J_{ISCO}$ denote the angular momentum flux and SAM at the ISCO, respectively. This treatment comes from the fact that the angular momentum of matter in the disk is transported outwards due to the turbulent viscosity or non-axisymmetric waves, and finally the matter falls into a black hole with $\sim J_{ISCO}$.

According to \citet{2011MNRAS.410.2302Z}, the jet luminosity by the neutrino process is determined by;
\begin{eqnarray}
&L_{j} = 1.1 \times 10^{52} x_{ms}^{-4.8}( \frac{M_{bh}}{3 M_{\odot}} )^{-\frac{3}{2}} \nonumber \\
& \hspace{8mm} \times  \begin{cases} 0 & (\dot{M} < \dot{M}_{ign}) \\
                 \dot{m}^{\frac{9}{4}} & ( \dot{M}_{ign} < \dot{M} < \dot{M}_{trap}) \\
                 \dot{m}_{trap}^{\frac{9}{4}} & (\dot{M} > \dot{M}_{trap}) \label{eq:neutrinolumi}
   \end{cases}
\end{eqnarray}
where $\dot{m} \equiv \dot{M}/(M_{\odot}/s)$, $x_{ms} \equiv r_{ms}/(2GM_{bh}/c^2)$ ($r_{ms}$ denotes the marginally stable orbit). $G$ and $c$ denote the gravitational constant and the speed of light, respectively. The characteristic mass accretion rate $\dot{M}_{ign}$ and $\dot{M}_{trap}$ are given as a function of the viscous parameter ($\alpha$), and the Kerr parameter ($a$) (see \citet{2011MNRAS.410.2302Z}).
Throughout this paper, we set $\alpha=0.1$ and Kerr parameter dependence on these accretion rate is linearly interpolated by $x_{ms}$.

%%%%%%%%%%%%%%%%%%%%%%%%

\begin{deluxetable*}{lccccccccccccc}
\tabletypesize{\scriptsize}
%\rotate
\tablecaption{Summary of our results \label{tab:model}} 
\tablewidth{0pt}
\startdata
\hline\hline
Model~~ &
~Breakout~  &
~$t_{i}$~ &
~$t_{br}$~ &
~$L_{p}$~ &
~$E_{dg}$~ &
~$E_{j}$~ &
~$E_{j>L_{50}}$~ &
~$E_{j>L_{49.5}}$~ &
~$T_{j>L_{50}}$~ &
~$T_{j>L_{49.5}}$~ &\\
      &     &  (s) & (s)  & ($10^{50}$ erg/s) & ($10^{51}$erg)& ($10^{51}$erg) & ($10^{51}$erg) & ($10^{51}$erg) & (s) & (s) \\
\hline
Mref  & yes & 10.9 & 27.8 & 1.9 & 1.4 & 7.4  & 4.4 & 5.6 & 27.8 & 46.2 \\
M150  & yes & 2.2  & 17.6 & 3.2 & 1.6 & 11.7 & 8.8 & 9.8 & 40.0 & 55.0 \\
M70   & yes & 15.5 & 45.5 & 1.0 & 0.9 & 3.6  &  -  & 1.7 &  -   & 21.7 \\
M50   & no  & 21.6 &  -   & 0.6 &  -  &  -   &  -  &  -  &  -   &  -   \\
\enddata
%\tablecomments{}
\end{deluxetable*}
%%%%%%%%%%%%%%%%%%%%%

The spherical symmetric density distribution is mapped by the spherical coordinate. The computational domain covers from $10^{8}$ to $4 \times 10^{10}$cm. Note that the location of the inner boundary in the present study is ten times smaller than in previous jet propagation studies (see e.g. \cite{2009ApJ...700L..47L,2011ApJ...732...26M,2011ApJ...731...80N}). The evolution of mass accretion rate, which is sensitive to the location of inner boundary (see \cite{2012ApJ...754...85N}), can thus be better captured by our simulations. However, as a result, these simulations become rather computationally expensive and we are only able to conduct them until the jet bow shock reaches the stellar surface or the black hole mass reaches $10 M_{\odot}$. The evolution of the the post-breakout phase is then analyzed by using a simple analytic formalism (see Eqs.~(\ref{eq:timeextrapo})-(\ref{eq:acrateana})). The results of an extended numerical simulation will be nonetheless compare with the analytical approach for the reference model in order to confirm that the analytical approach qualitatively captures the evolution of the jet dynamics (see Section~\ref{sec:subsecpostbreak} and Figure~\ref{f4}).

 We employ the gamma-law equation of state with $\gamma = 4/3$. The jet injection parameters such as the Lorentz factor and the specific internal energy are the same as those used in the standard model of \citet{2012ApJ...754...85N}, where the initial Lorentz factor and specific internal energy are fixed to $\Gamma=400$ and $\epsilon=0.01$, respectively. It should be noted for a fixed $\Gamma$ and $\epsilon$, the overall jet dynamics depend solely on $\theta_{op}$ (see \citet{2012ApJ...754...85N}). In this study, we assume $\theta_{op} = 9^{\circ}$, which agrees well with the opening angles deduced for Long GRBs \citep{2011arXiv1101.2458G}. The dependence of our results on the $\theta_{op}$ will be discussed in section~\ref{sec:results}.

 The 1000 non-uniform radial grids cover all the computational region while the meridian section is covered by 60 uniform grids. The 3 level Adaptive Mesh Refinement technique, similar to that used in \citet{2011ApJ...731...80N,2012ApJ...754...85N} is also employed in order to decrease computational cost.
We set up the stellar rotation by a similar manner as those used in \citet{2010ApJ...716.1308L}. The SAM distribution is separated into radial and polar components as $J(r,\theta) = j(r) \Theta(\theta)$, where $r$ and $\theta$ are the spherical radius and polar angle, respectively. In the reference model (Mref), $j(r)$ is given by 16TI model. For M150, M70 and M50 models, $j(r)$ is multiplied by 1.5, 0.7 and 0.5, respectively (see Figure~\ref{f1} for the SAM distribution of our models). The polar angle components are assumed to be rigid body rotation on shells, i.e., $\Theta(\theta) = \sin^2{\theta}$.

It is important to highlight that the simulations in this study do not cover the black hole accretion disk system. Even if the simulations cover the full computational domain, our numerical calculations can not treat the disk evolution appropriately, since the general relativistic effect and microphysics, which are important in determining the disk evolution, are not incorporated. However, the analytical formula proposed by \citet{2011MNRAS.410.2302Z} allows us to estimate neutrino luminosity without resolving the black hole accretion disk system. Owing to these prescription, we can study the jet penetration phase as determined by the neutrino-driven energy injection. The study of the coupling between black hole and disk accretion system is the beyond the scope of this paper.

We also note that the injection of the jet is delayed in the simulation as a result of the inner core not possessing enough angular momentum to create a disk \citep{2006ApJ...637..914W,2006ApJ...641..961L,2011MNRAS.410.2302Z}. Based on the standard neutrino-driven collapsar model (and assumptions those used in \citet{2007ApJ...657..383C,2011MNRAS.410.2302Z}), the central engine starts to operate after the accretion disk is formed around a black hole. Therefore, in our simulations, the jet is injected only when the SAM of matter at the inner boundary in equatorial plane exceeds $J_{ISCO}$.

\section{Results}
\label{sec:results}

\subsection{Basic features and jet penetrability}
\label{sec:subsecbasic}

\begin{figure}
\vspace{15mm}
\epsscale{1.2}
\plotone{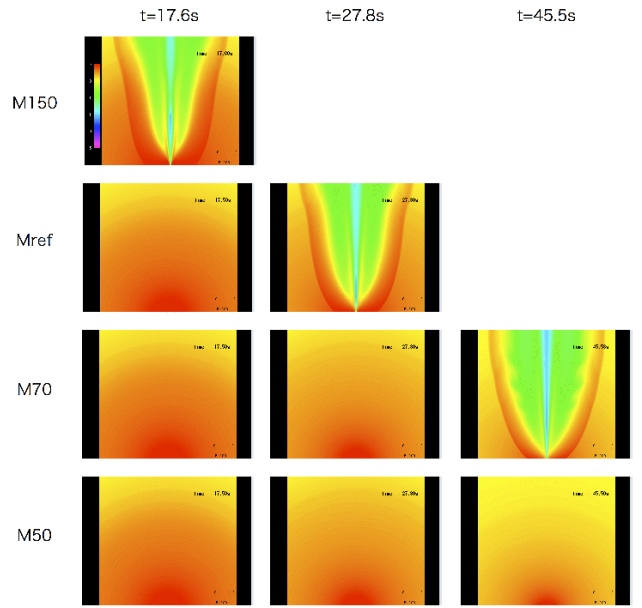}
\caption{The density contour (log scale) in the meridian section for each model at $t=17.6$s (left), $t=27.8$s (middle) and $t=46.5$s (right). Each time corresponds to the time of jet breakout for M150, Mref and M70, respectively (See also $t_{br}$ in Table~\ref{tab:model}). The spatial size of each box is $-5 \times 10^{9}$cm $< x < 5 \times 10^{9}$cm and $0 < z < 1 \times 10^{10}$cm. The upper and lower panels correspond to M150, Mref, M70 and M50 models. Note that some panels in M150 and Mref are lacking since we stop the simulation at the time of breakout.
\label{f2}}
\end{figure}

The overall evolution of the collapse of the progenitor seen in our simulations is not surprisingly similar to that found in \citet{2011ApJ...731...80N,2012ApJ...754...85N}. The infall of stellar envelope generates a rarefaction wave, which propagates outwards. The envelope contraction is almost identical among all models and follows a rather spherical contraction, since the centrifugal force plays a minor role. Note that we find that the density distribution at the inner boundary is slightly oblate but this does not affect the subsequent jet propagation although it might have consequences for the jet production (which is not properly simulated here).

\begin{figure}
\vspace{15mm}
\epsscale{1.1}
\plotone{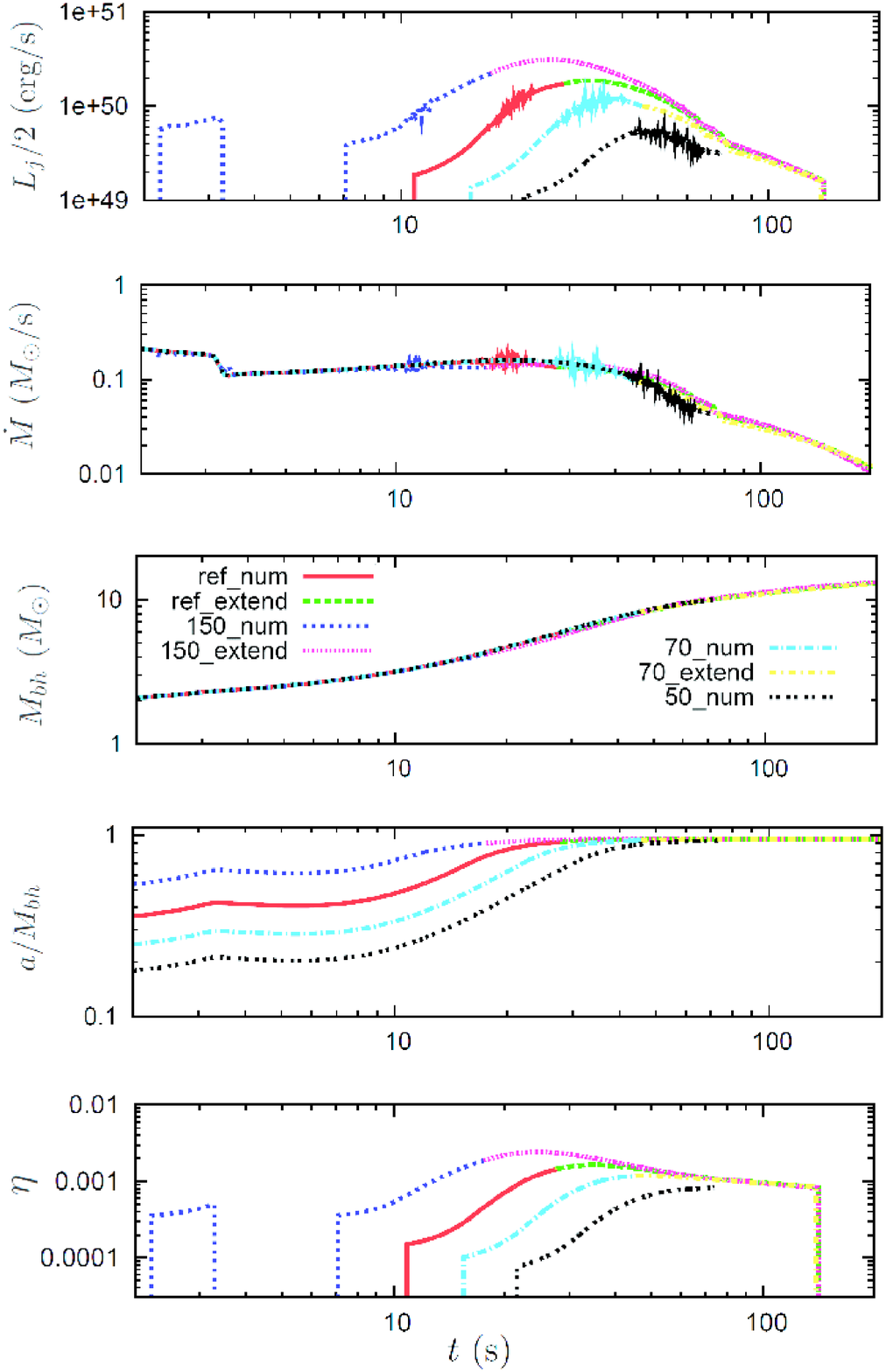}
\caption{From the top to the bottom, hemispherical neutrino luminosity, mass accretion rate, black hole mass, spin parameter and conversion efficiency from accretion energy to neutrino luminosity ($\eta \equiv L_{j}/\dot{M} c^2$) as a function of time from the onset of the collapse. Each color represents different models. Note that the model name with ``extend''  indicates the extrapolate results by analytic treatment (see text for more details.).
\label{f3}}
\end{figure}

 During the jet propagation phase, on the other hand, the results of jet evolution are very different among each model (see Table~\ref{tab:model} and Figure~\ref{f2}). We also show that the summary of our results in Table~\ref{tab:model}. For model M50, the forward shock wave does not move out and almost stagnates around the inner boundary despite the successful operation of the central engine ($t_{i}$ in Table~\ref{tab:model} denotes the time of initiation of central engine). In fact, no collimated feature can be seen for M50 in the lowest panels in Figure~\ref{f2}. This is attributed to the fact that the jet power does not exceed the ram pressure of the inflowing material, and the forward shock wave stagnates or is advected inwards. For models with successful jet breakout, on the other hand, the jet also can not move forward quickly after the initiation of the central engine. However, due to an increase in the Kerr parameter of the black hole over time, the jet power eventually exceeds the ram pressure of the inflowing material (Figure~\ref{f3}). Once the forward shock wave is able to move out, the jet interacts with the stellar mantle and gives rise to a cocoon. The hot cocoon helps jet confinement and helps to preserve the jet's strong outgoing momentum and energy flux, eventually leading to a successful breakout.

Figure~\ref{f3} shows the evolution of hemispherical neutrino luminosity, mass accretion rate, black hole mass, Kerr parameter and conversion efficiency from accretion energy to neutrino luminosity ($\eta \equiv L_{j}/\dot{M} c^2$) as a function of time from the onset of the collapse. The chief cause for the different jet propagation behavior is the sensitive dependence of the neutrino luminosity on the Kerr parameter (see Eq.~(\ref{eq:neutrinolumi})). For the fast rotational model, the angular momentum of the black hole is very large and it increases with time (see the 4th panel in Figure~\ref{f3}), which produces a powerful jet as a result of the large neutrino deposition energy rates. It is also important to note that the onset timing of central engine ($t_i$) also greatly affects the outcome of explosion. As shown in Table~\ref{tab:model} and illustrated in Figure~\ref{f2}, the jet production is significantly delayed for a slower rotational model. This is due to the neutrino luminosity being weaker for both smaller accretion rates and larger black hole masses (see Eq.~(\ref{eq:neutrinolumi}) for the dependence of $\dot{m}$ and $M_{bh}$). In fact, for M50, although the Kerr parameter reaches $\gtrsim 0.9$ at the end of our simulation, the neutrino luminosity is not large enough to move out the forward shock wave.

 According to these results, we infer that neutrino driven jets may not penetrate progenitors with extended envelopes, since significant large mass might be able to accrete to the black hole before jet breakout. The weak neutrino luminosity resulting for a sizable increase in black hole mass could result in the jet becoming non-relativistic ejecta. Therefore a compact progenitor is an inevitable requirement for a successful of neutrino-driven jet breakout. For the massive envelope progenitors such as PoP III or Red (Blue) supergiants, a different process might be required to powered GRBs (see also discussions in \citet{2011ApJ...726..107S}).

The foremost important result in this study is that the jet succeeds to break out from the star except for M50, which corresponds to the model with the slowest rotation rate. The neutrino-driven jet with $\theta_{op}=9^{\circ}$ produced in a rapidly rotating compact Wolf-Rayet star can potentially give rise to a GRBs. We also find that the time-averaged accretion-to-jet conversion efficiency among successful jet breakout models is roughly $\eta \sim 10^{-3}$, while $\eta$ for M50 can not reach $10^{-3}$ and never accomplish the jet breakout. This result is roughly consistent with our previous work \citep{2012ApJ...754...85N}. The analytical criteria in \citep{2012ApJ...754...85N} also shows the opening angle dependence for the successful jet breakout, which is the threshold $\eta$ increases with $(\theta_{op})^2$. Therefore, according to this result, we would like to give an important caution that jets with wider opening angle are more difficult to penetrate the star than the present results, which indicates that the threshold progenitor rotation rate differs in accordance with the jet opening angle.

\subsection{Post breakout phase: Analytical Formula}
\label{sec:subsecpostbreak}
As we have already mentioned, our numerical simulations are terminated at the time of jet breakout. However, it is interesting to extend the result of our numerical simulations to post break out stage, which allows us to estimate the expected observational differences among the computed models (see Section~\ref{sec:subsecobser}).
 We employee the following analytic approximations in order to see the subsequent evolution after the jet break out;
\begin{eqnarray}
t &=& t_{(b)} + \int_{r_{(b)}}^{r} \frac{dt_{ff}}{dr} (r) dr \label{eq:timeextrapo}\\ 
M_{bh} &=& M_{bh(b)} + \int_{r_{(b)}}^{r} \frac{dM_{r}}{dr} (r) dr \label{eq:massextrapo}
\end{eqnarray}
where
\begin{eqnarray}
t_{ff}(r) &=& \beta \sqrt{\frac{GM_{r}}{r^3}} \label{eq:freefallana}, \\
\dot{M}_{ff}(r) &=& \frac{dM_{r}/dr}{dt_{ff}/dr} \nonumber \\
 &=& \frac{1}{\beta ^2}
   \frac{8 \pi G M_{r} t_{ff} \rho }
               {3 M_{r} - 4 \pi r^3 \rho } \label{eq:acrateana}.
\end{eqnarray}
The above analytical estimation is essentially similar to the approach presented in \citet{2012ApJ...754...85N,2011ApJ...726..107S}. The free fall time $t_{ff}(r)$ can be determined by based on the assumption of spherical symmetric envelope contraction.
 Here, $t_{b}$ denotes the time at jet break out. Note also that the functions of $M_{r}(r)$ and $\rho(r)$ are extracted from the table of 16TI model. $r_b$ is determined by the assumption of $M_{bh} (t_{b}) = M_r (r_{(b)})$. The non-dimensional parameter $\beta$ is determined to ensure that $\dot{M}(t_b)$ is equal to $\dot{M}_{ff}(r_{(b)})$. According to this procedure, the neutrino luminosity and mass accretion rate can be smoothly connected from the results of numerical simulations. The evolution of the Kerr parameter is determined by Eq.~(\ref{eq:anguinteg}). Note that, since model M50 does not succeed the jet breakout, there is no post breakout phase for this model. Results of these analytical extension are also described in Figure~\ref{f3}.
\begin{figure}
\vspace{15mm}
\epsscale{1.0}
\plotone{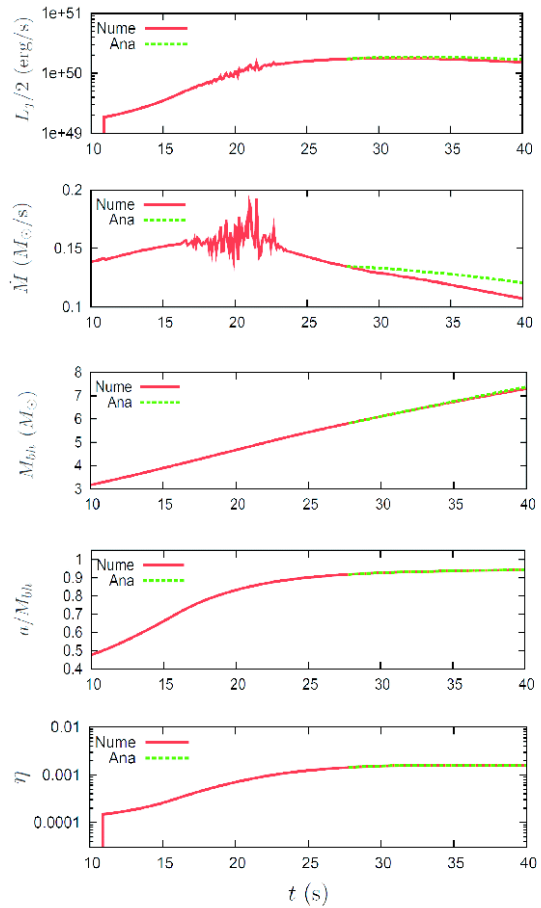}
\caption{Same as Figure~\ref{f3}, but comparison between results from extended numerical simulations and analytical approaches for the reference model.
\label{f4}}
\end{figure}
Figure~\ref{f4} shows the comparison between the results of extended numerical simulations and analytical estimation for the reference model. The extended numerical simulations are performed until $t=40$s, which corresponds to about ten seconds after the jet breakout. Note that we do not broaden the computational region, since the stellar contraction is not affected by the jet dynamics in the outer parts of the star. As shown in this figure, the time evolution of black hole mass ($M_{bh}$), Kerr parameter ($a$) and conversion efficiency ($\eta$) are almost identical between results of the extended numerical simulation and the analytical calculation. For luminosity ($L_{j}$) and mass accretion rate ($\dot{M}$), on the other hand, the analytical calculations are slightly larger than the results of our numerical simulations. This may be attributed to the fact that the analytical approach neglects stellar rotation, which increases the mass accretion rate, and consequently overestimates the neutrino luminosity. However, these differences are within ten percent. Therefore, we confirm that the above analytical approach qualitatively well describes the time evolution of the dynamics of the jet. In the following subsection, we discuss the observational consequence with the aid of the analytical approach in the post breakout phase.

\subsection{Observational consequences}
\label{sec:subsecobser}
 We first divide the energetics of the jet into two parts, which are relativistic jet component ($E_j$) and the cocoon component ($E_{dg}$) (see \citet{2011ApJ...726..107S,2003MNRAS.345..575M}). The $E_j$ is calculated based on the assumption that all the injected energy after the jet break out goes into the relativistic component, i.e, $E_j$ is given by $\int_{t_b}^{\infty} (L_{j}/2)  dt$. Note that $1/2$ factor comes from the assumption of equatorial symmetry.

\begin{figure}
\vspace{15mm}
\epsscale{1.2}
\plotone{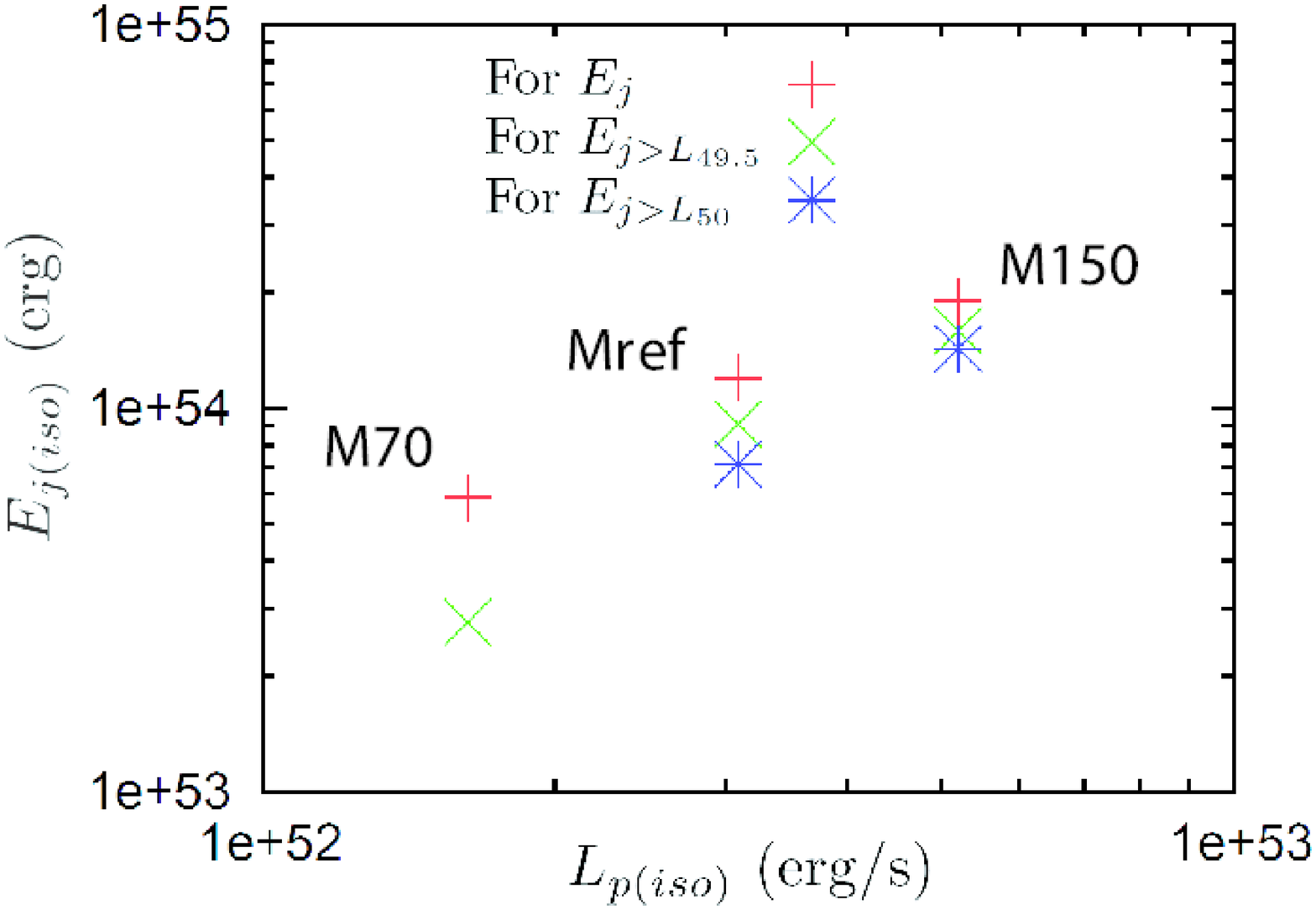}
\caption{$E_{j(iso)}$ and $L_{p(iso)}$ relation among models with the successful jet breakout. $E_{j(iso)}$ is calculated from $E_{j}$ (Red), $E_{j>L_{49.5}}$ (Green) and $E_{j>L_{50}}$ (Blue), respectively.
\label{f5}}
\end{figure}

As shown in Table~\ref{tab:model}, $E_j$ increases with an increase of stellar rotation. For the purpose of studying the outcome of explosion in more detail, we further divide the energy of relativistic jet into $E_{j>L_{50}}$ and $E_{j>L_{49.5}}$. $E_{j>L_{50}}$ is calculated by the same manner as $E_j$ except for the integration is carried out when $L_j/2 > 10^{50}$erg/s, while $E_{j>L_{49.5}}$ is calculated with a condition $L_j/2 > 5 \times 10^{49}$erg/s. In addition, we also list the corresponding time duration of each component as $T_{j>L_{50}}$ and $T_{j>L_{49.5}}$ in Table~\ref{tab:model}. In all cases, these timescales are several tens of seconds, which are comparable with typical time scale of prompt phase of GRBs. It should be noted, however, according to \citet{2012ApJ...749..110B}, the duration of prompt phase of GRBs may be modified by the duration of jet penetration, i.e, $t_{\gamma} = t_{e} - t_{b}$ (see also Eq.~(2) in \citet{2012ApJ...749..110B}), where $t_{\gamma}$, $t_{e}$, $t_{b}$ denote observed duration of the prompt phase, the central engine working time and the duration of the jet penetration phase, respectively. Therefore, the actual observed duration will be smaller than the $T_{j}$, and it is substantially modified especially for slower rotation models (e.g. M70) (since $t_{b}$ is larger with slower rotation).
 It is also interesting to note that model M70, which is the slowest model among successful jet breakout models, produces the weak explosion and does not have $E_{j>L_{50}}$ due to the low luminosity of the jet. Note also that this luminosity is upper limit for observed luminosity since we neglect the conversion process from hydrodynamical energy to gamma-rays. According to these facts, we infer that the neutrino-driven jet from the compact Wolf-Rayet star, whose rotation rate is between model M70 and M50, may produce very low luminous type burst, possibly LLGRBs. These results are qualitatively consistent with the previous studies (see e.g. \citet{2009ApJ...692..804L,2010ApJ...713..800L,2012ApJ...744..103M}). It should be noted, however, that some LLGRBs are observed with the extreme long duration, which populations can not be explained by the results of current studies. Therefore, the neutrino-driven jet may contribute to only LLGRBs with typical duration of prompt burst ($\sim 10$s).

On the other hand, the energy of cocoon component can be estimated as the diagnostic energy at $t=t_b$ (see \citet{2012ApJ...754...85N}). It is important to note that, as shown in Table~\ref{tab:model}, $E_{dg}$ is typically $\sim 10^{51}$erg for models with successful jet breakout.
This may be attributed the fact that the jet with a slowly rotating core tends to be weaker power and it spends longer time to penetrate the star. Therefore, despite its low jet luminosity, a large fraction of jet energy has been consumed for sweeping aside the stellar mantle and accumulated as the cocoon energy, which eventually reaches $\sim 10^{51}$erg. The cocoon material is expected to contribute the subsequent explosive event after the prompt phase \citep{2007ApJ...657L..77T,2012ApJ...750...68L} and at the afterglow phase \citep{2002MNRAS.337.1349R}.
The results of neutrino-deposition presented in this paper are not able to discern whether sufficient large nickel production might take place to explain hypernova explosions (see also discussions of cocoon propagation in \citet{2001ApJ...556L..37M,2002MNRAS.337.1349R,2003MNRAS.345..575M}). If nickel is not effectively produce at the jet interaction region alternative pathways such as the disk wind by viscous-heating or magnetic-driven wind from central engine would be required to explain the link between the GRBs-Hypernovae.

We further calculate the isotropic energy ($E_{j(iso)}$) for $E_{j}$, $E_{j>L_{49.5}}$ and $E_{j>L_{50}}$, and also isotropic peak luminosity $L_{p(iso)}$, which are shown in Figure~\ref{f5}. In this calculation, the jet opening angle is assumed to be $\theta_{op}= 9^{\circ}$, which is the same as the root of injected jet in our simulations. Note again that we neglect the conversion efficiency from hydrodynamic energy to radiation, so our results are still at the qualitative level and give the only upper limit of GRB radiation. In addition, the time evolution of neutrino luminosity does not capture the rapid time variability of central engine since our numerical simulations do not incorporate the black hole accretion disk system. According to these ambiguities, the time evolution of neutrino luminosity as described in the upper panel of Figure~\ref{f3} are different from the observed light curve in reality. It should be noted, however, that our analysis is meaningful to give the constraint the neutrino luminosity and energy as the upper limit. Based on the above assumption, we find that $E_{j(iso)}$ is $\sim 10^{54}$erg, while $L_{p(iso)}$ is $\sim 10^{52}$erg/s, which are sufficiently large to explain GRBs. We would like to point out that, for the jet with for rapidly rotating progenitor (M150), large fraction of energy are radiated with high luminosity jet ($L_j > 10^{50}$erg/s), while more than half of jet energy for M70 would be radiated with low luminous jet ($L_j < 5 \times 10^{49}$erg/s). According to these results, we suggest that neutrino-driven jet is capable of producing several types of bursts by the different progenitor rotation, which may be the origin of observational different bursts such as GRBs, XRFs. The failed GRBs would be also explained by neutrino-driven central engine when the progenitor is slowly rotating.

\section{Summary}

We present the numerical results of neutrino-driven jet propagation in a rotating Wolf-Rayet star. By changing the rate of progenitor rotation, we discuss jet penetrability and their observational consequences with the aide of analytic extrapolation in the post breakout phase. We show that every model except for M50 succeeds to break out the star. Especially, Mref and M150, which correspond models with sufficient rapidly rotation, have the relativistic outflow component as $L_{p(iso)}\sim 10^{52}$erg/s and $E_{j(iso)}\sim 10^{54}$erg, which are sufficiently large to explain GRBs. On the other hand, the energy in the cocoon component $E_{dg}$ is $\sim 10^{51}$erg for models with successful jet break outs, although it remains an open question whether the jet or the cocoon expansion could give rise to enough nickel production to explain the GRBs-Hypernovae connection. One of the other important results in this study is that model M50, which corresponds the model with slowly rotational model, can not succeed the jet breakout (failed GRB). Therefore, there is the threshold SAM distribution between model M70 and M50 for the success of jet penetration. It should be noted, however, that the threshold rotation, no doubt, strongly depend on the jet opening angle, and our results are only adequate for the canonical jet opening angle, $\theta_{op} = 9^{\circ}$. Although there are some limitations in this study, we suggest that neutrino-driven jet is capable of producing several types of bursts (and also include failure branch of burst i.e, failed GRBs) by the different progenitor rotation.

Finally, we would like to note that the results presented in this paper are optimistic. As one of the large uncertainties, the actual mass accretion rate would be smaller than the results obtained in this paper, since some fraction of the mass is expected to escape from the disk rather than being accreted to the black hole due to neutrino winds or viscous heating \citep{1999ApJ...524..262M}. In addition, the increase rate of Kerr parameter would be slower than the current result, since the SAM of the inflow matter is smaller than the ISCO due to the effect of pressure gradient. Note also that the disk wind also extract the angular momentum of accretion matter. Since the neutrino deposition rate depends sensitively on the mass accretion rate and Kerr parameter, the jet dynamics would be affected by these effects. Note also that the neutrino-driven jet can not explain the extreme long duration of bursts and other populations are necessary to explain these peculiar events. The other factors such as viewing angle may also cause the observational difference among GRB population (See e.g. \citet{2002ApJ...571L..31Y,2003ApJ...594L..79Y,2005ApJ...630.1003G}). More quantitative discussions will be conducted in our forthcoming paper.

\acknowledgments 
H.N is grateful to  Andrew Macfadyen, Andrei Beloborodov, Philipp Podsiadlowski, Kunihito Ioka, Yudai Suwa and Shoichi Yamada for useful comments and discussions. H.N would also like to thank Eriko Nagakura for proofreadings. This work was supported by Grant-in-Aid for the Scientific Research from the Ministry of Education, Culture, Sports, Science and Technology (MEXT), Japan (24740165) and HPCI Strategic Program of Japanese MEXT.

\end{document}